\documentclass[a4paper, 10pt]{article}
\usepackage[margin=2cm]{geometry}

\usepackage{booktabs}
\usepackage{bm}
\usepackage{tikz}
\usetikzlibrary{quantikz2}
\usepackage[version=4]{mhchem}
\usepackage{authblk}

\newcommand{\initphase}{
\begin{quantikz}
        \lstick{$\ket{0_k}$} & \gate{H} & \gate{P(\theta_k)} & \qw
        \end{quantikz}}

\newcommand{\initham}{
    \begin{quantikz}
        \lstick{$\ket{0_0}$} & \gate[wires=4]{\mathrm{Ansatz}} & \gate{U_0} & \qw
        \\ \lstick{$\ket{0_1}$} &  & \gate{U_1} & \qw
        \\ \lstick{$\cdots$} &  & \gate{U_k} & \qw
        \\ \lstick{$\ket{0_N}$} &  & \gate{U_N} & \qw
    \end{quantikz}
}

\newcommand{\foracle}{
    \begin{quantikz}
       \lstick{$o_0$}     & \gate[wires=6]{\mathrm{Sub.}} & \qw      &  \qw &  \qw     &\gate[wires=6]{\mathrm{Adder} } & \qw
    \\ \lstick{$\cdots$}     &                \qw               & \qw      &  \qw &  \qw     &                \qw                & \qw
    \\ \lstick{$o_N$}     &                               & \qw      &  \qw &  \qw     &                                & \qw
    \\ \lstick{$f_0$}     &                               & \qw      &  \qw &  \qw     &                                & \qw
    \\ \lstick{$\cdots$}     &                \qw               & \qw      &  \qw &  \qw     &                \qw                & \qw
    \\ \lstick{$f_N$}     &                   & \gate{X} &  \ctrl{1}    & \gate{X} & \qw & \qw
    \\ \lstick{$\ket{0}$} &  \qw              &  \qw     &  \targ{}    & \qw       & \qw & \qw
    \end{quantikz}
    }

\newcommand{\crossovercnot}{
    \begin{quantikz}
       \lstick{$\ket{\psi_k}$}     & \gate[wires=2]{H \oplus H} & \ctrl{1} & \qw
    \\ \lstick{$\ket{\psi_{k+1}}$} &                              & \targ{} & \qw
    \end{quantikz}}

\newcommand{\crossoveruqcm}{
    \begin{quantikz}
       \lstick{$\ket{\psi_k}$}     & \gate[wires=2]{\mathrm{UQCM} }  & \qw
    \\ \lstick{$\ket{\psi_{k+1}}$} &                              & \qw
    \end{quantikz}}

\newcommand{\fsadaptive}{
    \begin{quantikz}
       \lstick{$\ket{\psi_0}$} & \gate{H}&\ctrl{3}     &\gate{H}& \gate{R_y(\theta_0)} & \meter{}
    \\   \lstick{$\ket{\psi_{1} }$} &\gate{H}& \ctrl{2} &\gate{H}& \gate{R_y(\theta_{1})} & \meter{} 
    \\   \lstick{$\cdots$} & \gate{H}&\ctrl{1}   &\gate{H}& \gate{R_y(\theta_k)} & \meter{}
    \\   \lstick{$\ket{\psi_N}$} & \qw     & \targ{}   &  \qw      & \gate{R_y(\theta_N)} &   \qw &
    \end{quantikz}
    }

\newcommand{\fsgrover}{
    \begin{quantikz}
       \lstick{$\ket{\psi_0}$}      & \gate{H} & \gate[wires=4]{ O } & \gate{H} & \gate{X} & \gate[wires=4]{ D } & \gate{X} & \gate{H} & \qw
    \\ \lstick{$\ket{\psi_{1} }$} & \gate{H} &                     & \gate{H} & \gate{X} &  & \gate{X} & \gate{H} & \qw
    \\ \lstick{$\cdots $} & \gate{H} &                     & \gate{H} & \gate{X} &  & \gate{X} & \gate{H} & \qw 
    \\ \lstick{$\ket{\psi_N}$} & \qw &  \qw  & \qw  &  \qw &  \qw &  \qw & \qw  &  \qw
    \end{quantikz}}

\newcommand{\fsbubblesort}{
    \begin{quantikz}
       \lstick{$r_k$}          & \ctrl{2}    &  \gate{Z}    & \qw
       \\ \lstick{$r_{k+1}$}   & \ctrl{1}    &     \gate{Z}      &\qw
       \\ \lstick{$\ket{a_m}$} & \gate{\mathrm{Compare} } & \ctrl{-2} &  \qw
    \end{quantikz}}

\newcommand{\fsgeneticsampling}{
    \begin{quantikz}
       \lstick{$\ket{\psi_k}$} & \gate[wires=4]{\mathrm{Phase\, Flip} } & \gate[wires=4]{\mathrm{Mirror} } & \qw
    \\ \lstick{$\ket{\psi_{k+1} }$} &  &   & \qw
    \\ \lstick{$\cdots$}   &  &   & \qw
    \\ \lstick{$\ket{\psi_{N} }$} &  &   & \qw
    \end{quantikz}}

\newcommand{\pmutation}{
    \begin{quantikz}
    \lstick{$\ket{\psi_k}$} & \gate{P(\theta_k)} & \qw
    \end{quantikz}}

\newcommand{\umutation}{
    \begin{quantikz}
    \lstick{$\ket{\psi_k}$} & \gate{U^{w_k}} & \qw
    \end{quantikz}}

\newcommand{\cnotmutation}{
    \begin{quantikz}
    \lstick{$\ket{\psi_a}$}         & \ctrl{3} & \qw
    \\ \lstick{$\ket{\psi_{b} }$} & \ctrl{2}  & \qw
    \\ \lstick{$\ket{\psi_{c} }$} & \targ{} & \qw
    \\ \lstick{$\ket{\psi_{d} }$} & \targ{} & \qw
    \end{quantikz}}

\usepackage{braket}

\usepackage[numbers]{natbib}

\def\tsc#1{\csdef{#1}{\textsc{\lowercase{#1}}\xspace}}
\tsc{WGM}
\tsc{QE}
\tsc{EP}
\tsc{PMS}
\tsc{BEC}
\tsc{DE}

\title{Advances in Quantum Genetic Algorithms}

\author[1]{D. Lima}
\author[1]{R. Saini}
\author[1]{S. Al-Kuwari}
\affil[1]{Qatar Center for Quantum Computing, College of Science and Engineering, Hamad Bin Khalifa University, Doha, Qatar}
\date{17 Dec 2025}

\begin{document}
\maketitle

\begin{abstract}
Quantum Genetic Algorithms (QGAs) are an emerging field of multivariate quantum optimization that emulate Darwinian evolution and natural selection, with vast applications in chemistry and engineering. The appropriate application of fitness functions and fitness selection are the problem-encoding step and the slowest step in designing QGAs for specific physical applications. In this paper, we provide a comprehensive review of these crucial steps. Our survey maps cases of quantum advantage, classifies and illustrates QGAs and their subroutines, and discusses the two main physical problems tackled by QGAs: potential energy minimization of particles on a sphere, and molecular eigensolving. We conclude that the encoding used by the Thomson problem is a decisive step toward the use of QGAs in a variety of physical applications, while Grover's search as a selection step in Reduced QGAs is the main driver of quantum speedup.
\\ \quad
\\ \textbf{Keywords:} quantum genetic algorithm, fitness function, quantum evolutionary algorithm 
\end{abstract}

\maketitle

\section{Introduction}

\begin{figure}[ht]
        \centering
        \includegraphics[width=0.4\linewidth]{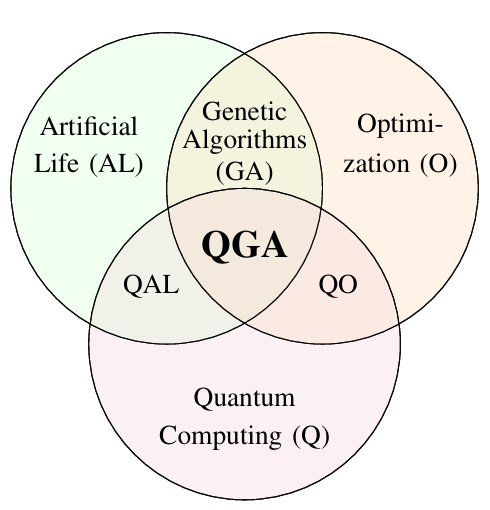}
        \caption{Venn diagram illustrates QGA as overlap of three different areas: Artificial Life, Optimization and Quantum Computing.}
        \label{fig:concept}
    \end{figure}

    Optimization techniques are widely used in industrial design, simulation and management to maximize quality and profit while minimizing errors. Recently, bio-inspired and quantum-improved algorithms have achieved high efficiency in optimization problems compared to their classical versions \cite{ibarrondo2022quantum}. The young field of Quantum Genetic Algorithms (QGA), i.e., evolutionary algorithms with an operation of crossover \cite{zhang2023quantum}, is an overlap of these areas (Fig. \ref{fig:concept}) that has matched expectations due to the high performance of its classical counterpart in the optimization of multi-variate systems \cite{ibarrondo2022quantum}.
     
    The implementation of a general Genetic Algorithm (GA) is constituted by a sequence of three operations over a set of data containers (e.g., strings or arrays): 
    \begin{itemize}
        \item[1)] crossover, the exchange of elements between two containers;
        \item[2)] mutation, the random transformation of elements in a random sample of all containers;
        \item[3)] fitness selection, the evaluation of a function over the elements of each container and subsequent rewarding of the containers in a good-fit range, or penalization of those outside the range.
    \end{itemize}
    
    The choice of vector as data type in a genetic algorithm is convenient, making it quantum-friendly \cite{amal2022quantum}, since a quantum algorithm is a sequence of unitary matrices (logical gates) used to manipulate probability amplitude vectors (also known as quantum states or state vectors). State vectors are formed by the tensor product of qubits (2D vectors), therefore $n$ qubits represent $2^n$ probable states \cite{zhang2021design}.
    
    The final step of every quantum algorithm, i.e., the measurement, is the sampling of a state from a distribution of states with different probabilities. The problem of maximizing the probability of one of these states arises naturally as a necessary step in the most relevant quantum algorithms to obtain the optimal value in a single run \cite{suo2020quantum}. Among the many available quantum optimization approaches, a Quantum Genetic Algorithm is a quantum computer-compatible adaptation of a classical GA, or a brand new model, with the advantage of being able to evaluate several properties of the containers simultaneously (rather than iteratively) during the fitness selection \cite{ibarrondo2022quantum}. It is also known to be easily adapted and combined with other algorithms, such as the pigeon swarm \cite{wang2023pid} and the immune algorithm \cite{bichara2023quantum}.
 
    In QGAs, qubits play the role of individuals, and the population is constituted by the quantum register, i.e., the set of all qubits. Qubits are sometimes referred to as chromosomes or genotypes, whereas the superposed states of a single qubit are its phenotypes \cite{banerjee2023robust}.

     \begin{figure*}[h]
         \centering
         \includegraphics[width=1\linewidth]{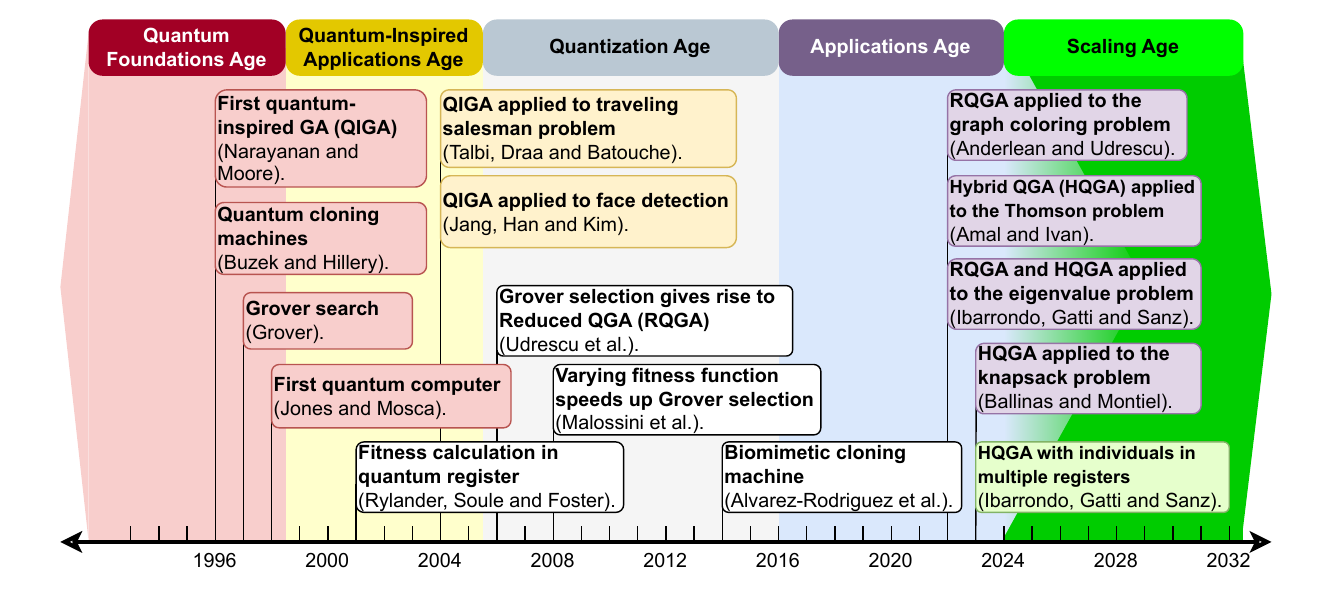}
         \caption{Timeline of QGA research from foundations of quantum computing to recent applications. References from top to bottom and left to right: (red boxes) \cite{narayanan1996quantum, buvzek1996quantum, grover1997quantum, jones1998implementation},
         (yellow boxes)
         \cite{talbi2004new, jang2004face},
         (white boxes) \cite{udrescu2006implementing, malossini2008quantum, rylander2001quantum, alvarez2014biomimetic},
         (purple boxes)
         \cite{ardelean2022graph, ibarrondo2022quantum, amal2022quantum, ballinas2023hybrid}, (green box) \cite{ibarrondo2023quantum}.
         }
         \label{fig:timechart}
     \end{figure*}

    \subsection{Applications of QGA}   %

        Research in QGA is marked by five periods (Fig. \ref{fig:timechart}), from the foundations of quantum algorithms and Quantum-Inspired Genetic Algorithms (QIGA) in 1996 to applications of QIGA after 1999, followed by a focus on adapting the standard GA steps to quantum computers in hybrid algorithms between 2006 and 2016, and finally, the contemporaneous period, marked by an increased number of publications in applications of hybrid QGAs after 2016, the year of the first peer-review survey in QGAs \cite{lahoz2016quantum}. Right after the Applications Age, the focus on scaling and partitioning prototypical applications to accommodate larger circuits and larger systems \cite{ibarrondo2023quantum} marks a smooth transition to the Scaling Age.  The milestones and main authors in the field are listed in Fig. \ref{fig:timechart}.
    
        The applications of QGAs comprise engineering problems, smart cities, cloud computing and text analysis. For example, synchronous motors that are used in blenders and other rotating engines are optimized by Wang \emph{et al.} \cite{wang2023pid} using QGA, considering three parameters of a PID speed controller for permanent magnet synchronous motors. These parameters are usually only obtained by heuristic experiments, subjecting the speed control of the motors to instability. The authors succeeded in showing that a custom, improved QGA can be used to find the minimum value of four special functions with a better performance than the previous classical and quantum GA, and thus achieve faster speed and stable state. 
        
        In another example, the design of IoT antennas is optimized using a QGA to reduce the antenna size by 87\% while maintaining a good radiation efficiency \cite{bichara2023quantum}, a relevant feature for portable devices such as smart bands and smartphones. 
        
        The rational approximation of wavelet base for wavelet filter in analog domain can also be optimized by QGA \cite{zhang2021design}. The installation of a grinding disk requires balanced control of the apparatus in order to avoid instability and to extend its lifespan. Zhang \emph{et al.} applied a QGA to efficiently optimize the balance \cite{zhang2022combination}. The design optimization of office building envelope using QGA results in 50\% shorter convergence time than the equivalent CGA \cite{wang2021design}. 
        
        The suitability of QGAs for optimization in combinatorial space makes it a good fit for the Salesman Problem. A very similar problem is that of schedule and route planning for buses in a large city \cite{lin2021intelligent}. The QGA model was able to lower the use of memory space, reduce the difficulty of generating the initial population and improve the search efficiency. The design of accessible emergency routes and shelter allocation is a multivariate problem in city planning that requires the minimization of costs, maximization of people inflow in case of an emergency, all under the limitation of the capacity of the evacuation site. This landscape is suitable for the use of QGA \cite{yin2023emergency} for higher efficiency than a classical genetic algorithm. A similar context is the task allocation problem, a problem that has a two-dimensional representation. Again, the best fit of the QGA converged to the optimum value in fewer generations than the two-dimensional classical GA used for comparison \cite{mondal2021two}.
     
        In another example, detection of plagiarism is a ubiquitous need for evaluation processes and publishing any sort of content. A semantic textual similarity estimation for plagiarism detection was designed with a QGA as a tool to accurately discriminate between original production and plagiarism \cite{darwish2023quantum}. In this case, accuracy is a higher priority than time efficiency due to the legal risks related with plagiarism. 
        
        Digital tools like cloud computing have a collection of combinatorial problems that can be solved by QGA, like scheduling in the fog-cloud computing environment, and the minimization of the execution time of a workflow. The authors addressed these topics with a new makespan quantum algorithm based on a classical workload task scheduling algorithm, obtaining the expected faster performance than the classical procedure \cite{belmahdi2022sqga}. As a complement, the optimization of multiple-input multiple output system (MIMO) was focused on energy efficiency. The simulations show that the suggested QGA efficiently determined the best transmit power of the active users, resulting in maximum energy efficiency \cite{almasaoodi2023optimizing}. 

\begin{figure*}
            \centering
            \includegraphics[width=0.7\linewidth]{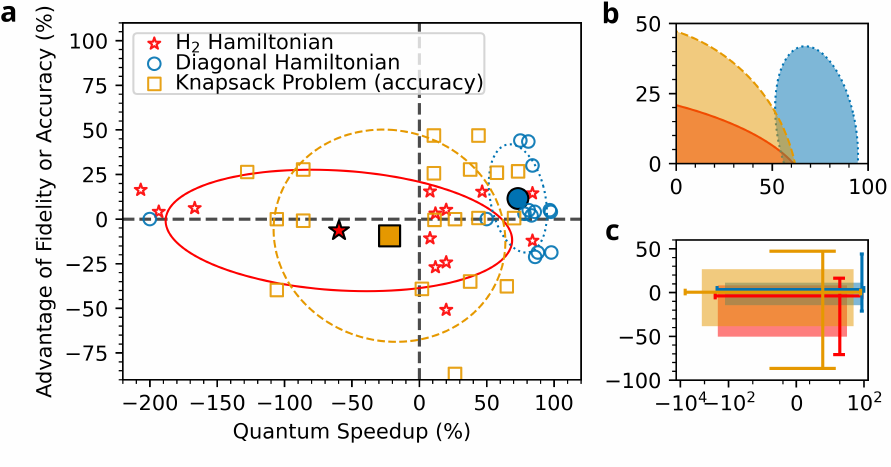}
            \caption{Quantum advantage of speedup versus fidelity (diagonal Hamiltonian and $\mathrm{H}_2$ Hamiltonian eigensolving problems) or accuracy (Knapsack Problem), adapted from \cite{ballinas2023hybrid} and \cite{ibarrondo2022quantum}, as compared with classical GAs. (a) Ellipses are Gaussian fittings after outlier removal. (b) Zoom-in of double advantage regions of the elliptical clusters. (c) Box plots including outliers. In the Knapsack Problem, six models of QGA were studied, and the populations had 10, 15 and 20 individuals. In the other two problems, four QGA models were studied, and the populations were all of four individuals.}
            \label{fig:adv}
        \end{figure*}
        
        The problem of tracking moving objects in videos is memory-intensive in face of high-quality and many-frame videos. As noted in the ease of adapting QGAs to different contexts, Zhang \emph{et al.} implemented a QGA assisted by Deep Learning on the fitness selection process to address that problem \cite{zhang2023quantum}. Their model was robust to environmental noise and changes in motion states, the relevant properties for this context.
 
        Fuzzy-based GAs share several properties with regular quantum genetic algorithms, such as emulating superposition and entanglement. For this reason, it was used to optimize a fog-based multimedia transmission schedule in the context of the Internet of Multimedia Things. The model was faster and more accurate than the other famous classical optimization routines, in addition to privilege diversity during the algorithm iteration \cite{zanbouri2023new}.

    \subsection{Quantum Advantages} \label{sec:advantages}
        Classical genetic algorithms that implement stochastic operations to emulate quantum gates are known as quantum-inspired genetic algorithms \cite{narayanan1996quantum, han2000genetic, xiong2018quantum}. Their goal is to approach quantum speedup through stochastic speedup \cite{han2000genetic, xiong2018quantum}, thus demonstrating problems that are suitable candidates for full adaptation to quantum circuits. The double digest problem is a representative example, since it involves concepts of molecular dynamics. The problem is to determine orders and distances between restriction sites on a DNA molecule by digesting the DNA with two restriction enzymes. The authors in \cite{suo2020quantum} claimed that this problem can be handled by a quantum-inspired GA more efficiently than a classical one once a sufficiently large quantum computer becomes accessible. Another notorious target for QGA is the density-based data clustering process. The authors used Euclidean proximity matrices as fitness score, and comprised crossover, mutation and selection in a single rotation operator. The algorithm is robust, with efficiency that grows directly proportional to the space \cite{banerjee2023robust}.
        
        A deeper model emulates entanglement in differential evolution algorithms. For example, in \cite{dixit2023quantum}, the crossover is probabilistic with a $0.9$ occurrence rate, and mutation is implemented via differential evolutionary perturbation. Such classical evolutionary algorithms can be enhanced by using quantum selection operators, although the improvement in performance may be small. In the model presented in \cite{von2022quantum}, only in 66\% of the cases was the quantum-enhanced algorithm faster than the classical counterpart. Their approach focused on maximizing the diversity score, in addition to the fitness score, by selecting parents that maintain both good fitness and good diversity scores. 
        
        Parametric comparative studies of QGAs and classical GAs were conducted by \cite{ballinas2023hybrid} and \cite{ibarrondo2022quantum}. Their results are cast in Fig. \ref{fig:adv}-a, showing the percent advantage of each QGA against the average convergence time and average fidelity of classical GA models for the Eigenvalue Problem of the \ce{H2} Hamiltonian and of a diagonal Hamiltonian, or accuracy in the case of the Knapsack Problem. The eigenvalue problem for the \ce{H2} and the diagonal Hamiltonians in Fig. \ref{fig:adv}-a encode 4 individuals in 8 qubits \cite{ibarrondo2022quantum}, while the Knapsack Problem \cite{ballinas2023hybrid} encodes 10, 15 and 20 individuals in less than 32 qubits, including ancilla qubits.
                
        As denoted by the double-advantage region in Fig. \ref{fig:adv}-b, while a quantum advantage of speedup is desirable, it is only reliable if accompanied by high fidelity. In addition, an inappropriate choice of parameters or a problem can make the QGA disadvantageous, as shown by the box plots in Fig. \ref{fig:adv}-c. 
        
        The \texttt{measure.EllipseModel} function of the \texttt{skimage} package in Python (version 3.13) programming language is used to define Gaussian elliptical clusters in Fig. \ref{fig:adv}-a , after outlier removal using a standard median-modified Z-score \cite{iglewicz1993volume} along the abscissa. The clusters are centered on the filled, larger marker of the same color. Fig. \ref{fig:adv}-b is a zoom-in of the first quadrant of Fig. \ref{fig:adv}-a, where the double-advantage regions of the clusters are shaded, showing that the fidelity and the accuracy advantages are topped at 50\%. 
        
        The box plot in Fig. \ref{fig:adv}-c included the whole dataset and shows that even when outliers are included, more than 50\% of the population in the three types of problem are still in the advantage region.

        Quantum emulation in classical computers suggests that quantum and hybrid genetic algorithms are not guaranteed to outperform classical ones for every application \cite{lahoz2016quantum}, although recent implementations in quantum processing units (QPUs) showed a speedup in hybrid models under problem-specific conditions \cite{ibarrondo2022quantum}. As any advantage in genetic algorithms is an emergent property, each step of the algorithm must be specifically designed and optimized for the type of problem to obtain maximum time and fidelity advantages.

    \subsection{Outline} %
        The rest of this paper is organized as follows: in section \ref{s:fundamentals} we introduce the fundamentals and main algorithms of Quantum Computing; in section \ref{s:qga}, we describe and exemplify the various implementations of core steps of QGA; in section \ref{s:implementations}, we classify the five most common QGA algorithms; in section \ref{s:tailoring}, we detail the main classes of fitness functions and describe applications of QGAs to problems of physical interest. We finalize with concluding remarks in section \ref{s:conclusion}.

    \section{Fundamentals of Quantum Computing} \label{s:fundamentals}
    Quantum computing operates on the principles of quantum mechanics to process information in ways that are fundamentally different from classical computation. At its core, a quantum computer encodes information in quantum states, or qubits. By harnessing phenomena such as superposition and entanglement, these machines can explore vast computational spaces and perform certain calculations exponentially faster than their classical counterparts \cite{nielsen2010quantum, imre2012advanced}. A quantum register of $n$ qubits can represent $2^n$ classical states simultaneously, providing a foundation for massive parallelism.

    A key principle of quantum computation is reversibility, which dictates that every computational step must be invertible. This is a direct consequence of the fact that quantum states evolve through unitary transformations, which are inherently reversible \cite{aruna2016study}. This constraint ensures that no information is lost during computation. To manage intermediate calculations, quantum algorithms often employ ancillary qubits, or ancillas, which are initialized to a known state and later discarded or measured. Ancillas are particularly crucial in quantum error correction, where they are used to detect errors in data qubits without disturbing the encoded information \cite{gottesman1997stabilizer}. The final step of any quantum computation is measurement, which projects the quantum state onto a classical outcome, yielding the result.

    The practical construction of a quantum computer is guided by the DiVincenzo criteria, which outline five essential requirements: a scalable system of well-characterized qubits, the ability to initialize qubits to a fiducial state, long coherence times to protect against environmental noise (decoherence), a universal set of quantum gates for arbitrary computations, and a high-fidelity measurement capability \cite{divincenzo2000physical}. Leading experimental platforms striving to meet these criteria include superconducting circuits, such as Google’s Sycamore processor \cite{arute2019quantum}, and trapped-ion systems, which have demonstrated high-fidelity multi-qubit operations \cite{wright2019benchmarking}.

\begin{figure}[ht]
      \centering
      \includegraphics[width=0.4\linewidth]{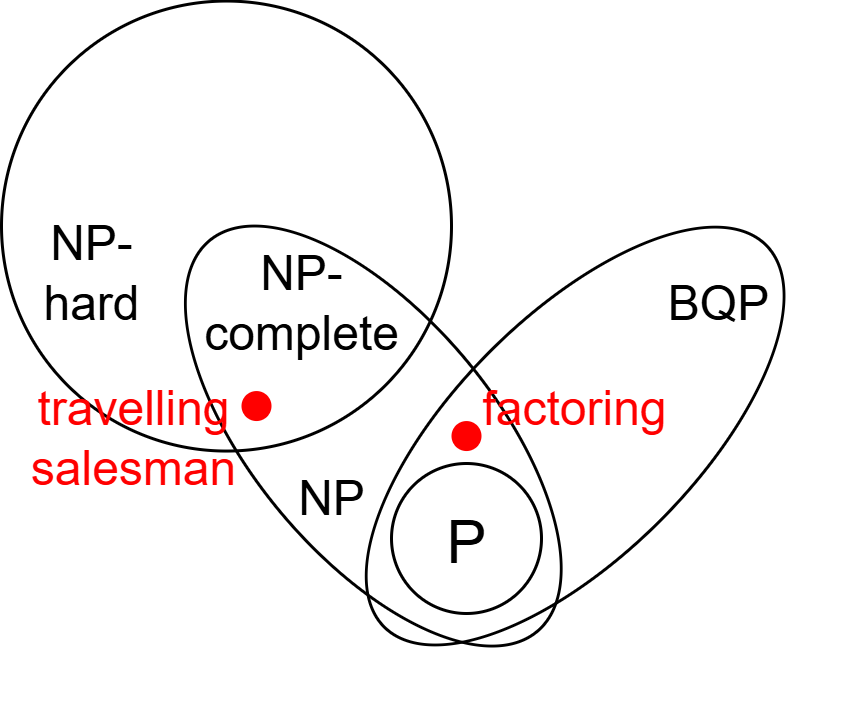}
      \caption{Venn diagram illustrating the conjectured relationships between classical complexity classes P, NP, and the quantum class BQP.}
      \label{fig:venn_complexity}
    \end{figure}

    The theoretical power of quantum computation is studied by quantum complexity theory, which explores the relationship between quantum complexity classes like BQP (Bounded-error Quantum Polynomial time) and classical classes such as P, NP, and the polynomial hierarchy (PH) \cite{aaronson2010bqp}, as illustrated in Fig. \ref{fig:venn_complexity}. A central finding in this field is that classical computers cannot efficiently simulate quantum systems, suggesting a fundamental separation in computational power \cite{aaronson2011computational, aaronson2013computational}. Furthermore, research on the "learnability" of quantum states has shown that they can be efficiently characterized through tomographic techniques, providing a framework for verifying quantum computations \cite{aaronson2007learnability}.

    \subsection{Qubits: The Foundation of Quantum Information}
    The fundamental unit of quantum information is the qubit, which can be realized physically using systems such as electron spins, photon polarizations, or atomic energy levels. Unlike a classical bit, which is restricted to a definite state of either 0 or 1, a qubit can exist in a superposition of both states simultaneously. This state, $\ket{\psi}$, is represented as a linear combination of the basis states $\ket{0}$ and $\ket{1}$:
    \begin{equation}
    \ket{\psi} = \alpha\ket{0} + \beta\ket{1}
    \label{psi}
    \end{equation}
    Here, $\alpha$ and $\beta$ are complex probability amplitudes that must satisfy the normalization condition $|\alpha|^2 + |\beta|^2 = 1$. This property enables a register of $n$ qubits to exist in a superposition of all $2^n$ possible classical states, providing the exponential computational space that underpins quantum advantage \cite{dirac1981principles}. For instance, Google's 53-qubit Sycamore processor can simultaneously represent $2^{53}$ (approximately  $9 \times 10^{15}$) states \cite{arute2019quantum}.
    \subsubsection{The Bloch Sphere Representation}
\begin{figure}[ht]
      \centering
      \includegraphics[width=0.4\linewidth]{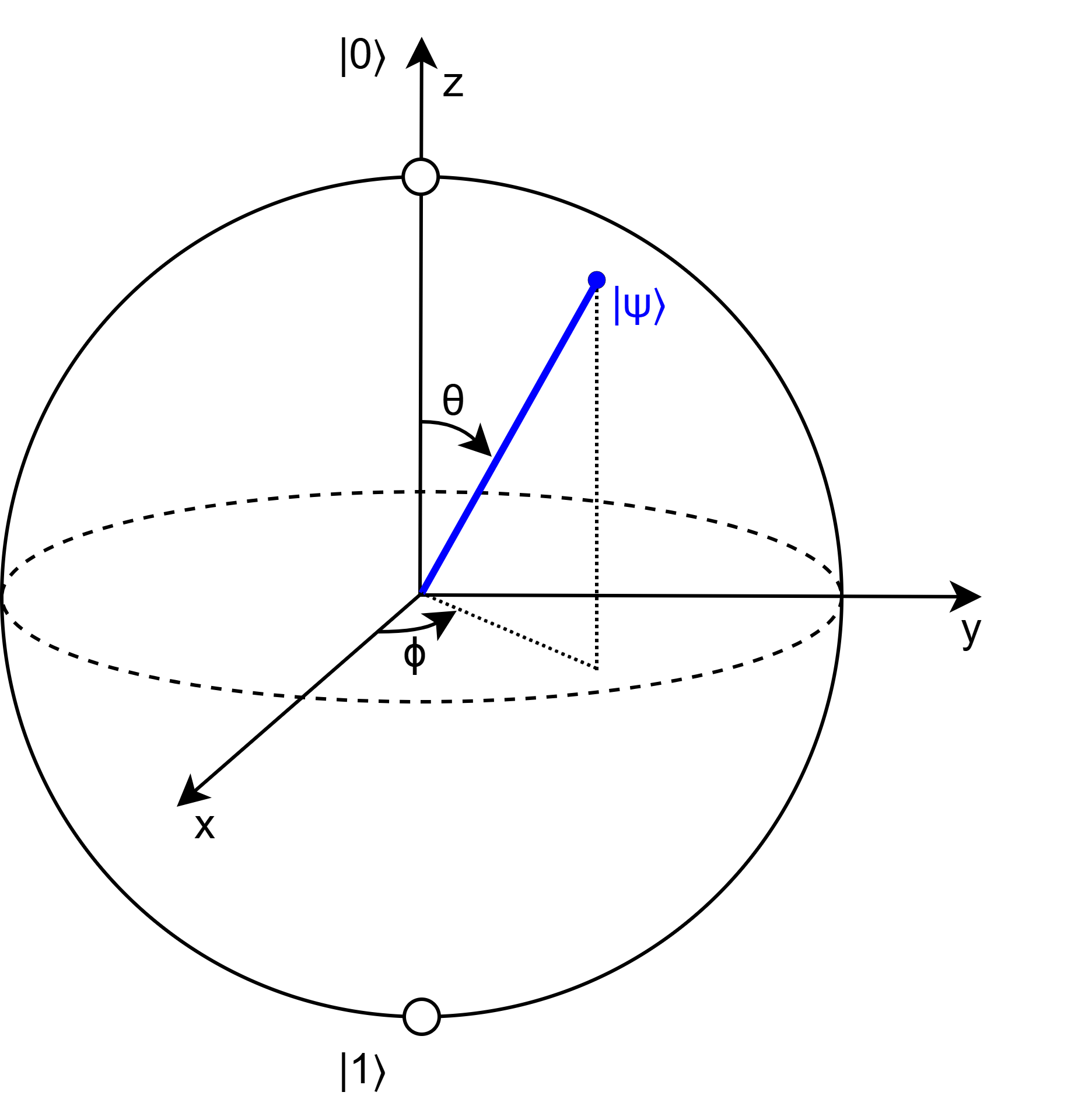}
      \caption{The Bloch sphere, a geometric representation of a single qubit state $|\psi\rangle$ defined by the polar angle $\theta$ and azimuthal angle $\phi$.}
      \label{fig:bloch_sphere}
    \end{figure}

    The state of a single qubit can be visualized geometrically using the Bloch sphere \cite{bloch1946nuclear}, as shown in Fig. \ref{fig:bloch_sphere}. In this sphere, the north and south poles correspond to the classical basis states $\ket{0}$ and $\ket{1}$, respectively. Any other point on the surface represents a unique superposition state. The state vector $\ket{\psi}$ is parameterized by two angles, $\theta$ (polar) and $\phi$ (azimuthal), leading to the general representation Eq. \ref{e:genrep}.
    \begin{equation} \label{e:genrep}
    \ket{\psi} = \cos\left(\frac{\theta}{2}\right)\ket{0} + \sin\left(\frac{\theta}{2}\right)e^{i\phi}\ket{1}
    \end{equation}
    where $0 \leq \theta \leq \pi$ and $0 \leq \phi < 2\pi$. The Bloch sphere provides an intuitive framework for understanding single-qubit operations, which correspond to rotations of the state vector on the sphere's surface.
    
    \subsection{Quantum Entanglement}
    While superposition describes the state of individual qubits, entanglement describes a uniquely quantum correlation between two or more qubits. When qubits are entangled, their fates are linked, regardless of the distance separating them. A measurement on one qubit instantaneously influences the state of the other(s). This phenomenon, famously dubbed "spooky action at a distance" by Albert Einstein, was a source of historical debate regarding the completeness of quantum mechanics (the EPR paradox) \cite{einstein1935can, laghari2022review}.
        
    The debate was largely settled by Bell's theorem, which proved that no classical theory of "local hidden variables" could reproduce the strong correlations predicted by quantum mechanics \cite{bell1964einstein, cacciapuoti2020entanglement}. Subsequent experiments, pioneered by Clauser, Aspect, and Zeilinger (Nobel Prize in Physics, 2022), have repeatedly violated Bell's inequality, confirming the non-local nature of entanglement. Today, entanglement is not just a theoretical curiosity but a critical resource for quantum algorithms, quantum communication, and quantum error correction \cite{horodecki2009quantum, erhard2020advances}.

    \subsection{Quantum Gates and Circuits}
    Quantum computations are implemented through sequences of quantum gates, which are reversible operations that manipulate the states of qubits. Mathematically, these gates are represented by unitary matrices. A set of single- and two-qubit gates is sufficient to construct any quantum computation, forming a universal gate set. A summary of common gates is provided in Table \ref{tab:single_two_qubit_gates}.

    Single-qubit gates perform rotations on the Bloch sphere. Key examples include:
    \begin{itemize}
        \item Pauli Gates (X, Y, Z): The X-gate acts as a quantum NOT, flipping $\ket{0} \Leftrightarrow \ket{1}$. The Z-gate applies a phase flip to the $\ket{1}$ state.
        \item Hadamard Gate (H): This crucial gate transforms basis states into equal superpositions, such as $\ket{0} \rightarrow \frac{1}{\sqrt{2}}(\ket{0} + \ket{1})$. Applying an H-gate to each qubit in an $n$-qubit register initialized to $\ket{0}^{\otimes n}$ creates a uniform superposition of all $2^n$ computational basis states.
    \end{itemize}
    Multi-qubit gates create correlations and entanglement between qubits. The most common is the Controlled-NOT (CNOT) gate, which flips the state of a target qubit if and only if a control qubit is in the state $\ket{1}$.

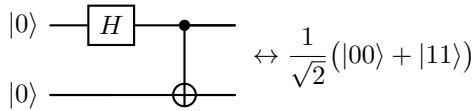
\begin{figure}[h]
  \centering
  \begin{quantikz}
    \lstick{$\ket{0}$} & \gate{H}      & \ctrl{1} & \qw \\
    \lstick{$\ket{0}$} & \qw           & \targ{}  & \qw 
  \end{quantikz}
  \vspace{6pt}
  $\displaystyle \leftrightarrow \frac{1}{\sqrt{2}}\bigl(\lvert00\rangle+\lvert11\rangle\bigr)$
  \caption{Quantum circuit to generate an entangled EPR pair (Bell state) from two qubits initialized to $\ket{00}$}
  \label{fig:epr_pair}
\end{figure}

    These gates are assembled into quantum circuits, which are analogous to classical logic circuits. For instance, the circuit in Fig. \ref{fig:epr_pair} uses a Hadamard gate followed by a CNOT gate to generate a maximally entangled Bell state or EPR pair: $\frac{1}{\sqrt{2}}(\ket{00} + \ket{11})$, demonstrating how fundamental gates produce non-classical states essential for quantum algorithms.

\begin{table*}[ht]
  \centering
  \caption{Common single- and two-qubit gates with their matrix representations and actions.}
  \label{tab:single_two_qubit_gates}
  \begin{tabular}{lcp{10cm}}
    \toprule
    Gate & Matrix representation & Gate action \\
    \midrule
    Identity $(I)$ &
      $\begin{pmatrix}1 & 0\\ 0 & 1\end{pmatrix}$ &
      Leaves the qubit state unchanged. \\[2ex]
    Pauli-$X$ &
      $\begin{pmatrix}0 & 1\\ 1 & 0\end{pmatrix}$ &
      Performs a bit-flip. \\[2ex]
    Pauli-$Y$ &
      $\begin{pmatrix}0 & -i\\ i & 0\end{pmatrix}$ &
      Performs a bit-flip combined with a phase-flip. \\[2ex]
    Pauli-$Z$ &
      $\begin{pmatrix}1 & 0\\ 0 & -1\end{pmatrix}$ &
      Applies a phase of $-1$ to the state $\ket{1}$. \\[2ex]
    Hadamard $(H)$ &
      $\frac{1}{\sqrt{2}}\begin{pmatrix}1 & 1\\ 1 & -1\end{pmatrix}$ &
      Creates an equal superposition of basis states. \\[2ex]
    Phase $(S)$ &
      $\begin{pmatrix}1 & 0\\ 0 & i\end{pmatrix}$ &
      Applies a phase of $i$ to the state $\ket{1}$. \\[2ex]
    $\pi/8$ Gate $(T)$ &
      $\begin{pmatrix}1 & 0\\ 0 & e^{i\pi/4}\end{pmatrix}$ &
      Applies a phase of $e^{i\pi/4}$ to the state $\ket{1}$. \\[2ex]
    $R_x(\theta)$ &
      $\begin{pmatrix}
        \cos\frac{\theta}{2} & -i\sin\frac{\theta}{2}\\
        -i\sin\frac{\theta}{2} & \cos\frac{\theta}{2}
      \end{pmatrix}$ &
      Rotation about the $X$-axis of the Bloch sphere by angle $\theta$. \\[3ex]
    $R_y(\theta)$ &
      $\begin{pmatrix}
        \cos\frac{\theta}{2} & -\sin\frac{\theta}{2}\\
        \sin\frac{\theta}{2} & \cos\frac{\theta}{2}
      \end{pmatrix}$ &
      Rotation about the $Y$-axis of the Bloch sphere by angle $\theta$. \\[3ex]
    $R_z(\theta)$ &
      $\begin{pmatrix}
        e^{-i\theta/2} & 0\\
        0 & e^{i\theta/2}
      \end{pmatrix}$ &
      Rotation about the $Z$-axis of the Bloch sphere by angle $\theta$. \\[3ex]
    CNOT &
      $\begin{pmatrix}
        1 & 0 & 0 & 0\\
        0 & 1 & 0 & 0\\
        0 & 0 & 0 & 1\\
        0 & 0 & 1 & 0
      \end{pmatrix}$ &
      Flips the target qubit if the control qubit is $\ket{1}$. \\[2ex]
    SWAP &
      $\begin{pmatrix}
        1 & 0 & 0 & 0\\
        0 & 0 & 1 & 0\\
        0 & 1 & 0 & 0\\
        0 & 0 & 0 & 1
      \end{pmatrix}$ &
      Swaps the quantum states of two qubits. \\[1.5ex]
    \bottomrule
  \end{tabular}
\end{table*}

\subsection{Quantum Algorithms}
Quantum algorithms are designed to exploit superposition and entanglement to solve specific problems more efficiently than any known classical algorithm. As shown in Table \ref{tab:algorithm_complexity}, the speedup they offer can be polynomial, superpolynomial, or exponential.

\subsubsection{Grover's Algorithm}
Developed by Lov Grover, this algorithm provides a quadratic speedup for searching an unsorted database \cite{grover1996fast}. While a classical search requires $O(N)$ queries on average to find an item among $N$ entries, Grover's algorithm achieves this in only $O(\sqrt{N})$ steps. It works by using an oracle to mark the target state and then systematically amplifying its probability amplitude through a process called "inversion about the mean." After approximately $\sqrt{N}$ iterations, a measurement of the quantum register will yield the target item with high probability.

\subsubsection{Shor's Algorithm}
Shor's algorithm demonstrates an exponential speedup for integer factorization, a problem considered intractable for classical computers for large numbers \cite{shor1999polynomial}. Its efficiency poses a significant threat to modern cryptography systems, such as RSA, which rely on the difficulty of factoring. The algorithm's core is a quantum subroutine for period-finding, which leverages the Quantum Fourier Transform (QFT). By preparing a superposition of states and applying the QFT, the algorithm efficiently finds the period of a modular exponentiation function, which can then be used to find the prime factors of the integer.

\subsubsection{Quantum Fourier Transform}
The QFT is the quantum analogue of the classical Discrete Fourier Transform (DFT) and is a key component in many quantum algorithms. It maps quantum states from the computational basis to the frequency (or Fourier) basis. Due to quantum parallelism, the QFT can be implemented on $n$ qubits with a complexity of $O(n^2)$ (equivalently $O(\log^2 N)$ for an $N = 2^n$ dimensional state space), which is an exponential improvement over the classical Fast Fourier Transform (FFT) complexity of $O(N \log N)$.

\begin{table*}[ht]
  \centering
  \caption{A summary of quantum versus classical algorithmic complexities for key problems.}
  \label{tab:algorithm_complexity}
  \begin{tabular}{llll}
    \hline
    Algorithm & Problem Type            & Quantum Complexity            & Classical Best       \\
    \hline
    Grover’s   & Unstructured search     & \(\mathcal{O}\left( \sqrt{N} \right)\)      & \(\mathcal{O}(N)\)    \\
    Shor’s     & Integer factorization   & Polynomial in \(\log N\)       & Sub‐exponential      \\
    QFT        & Fourier transform       & \(\mathcal{O}\left(\log^2 N \right)\)      & \(\mathcal{O}(N\log N)\) \\
    \hline
  \end{tabular}
\end{table*}

    \section{Quantum Genetic Algorithms} \label{s:qga}
        In this section, we describe and illustrate quantum gates of the main steps of a general quantum genetic algorithm.
        
        For a GA to be formally understood as quantum, and not a quantum-inspired GA, it is required to have a mapping from individuals to qubits and from the basic GA steps to quantum gates \cite{lahoz2016quantum}. In practice, the difference may be subtle, as most implementations of QGA are simulations in classical computers with some, but not all steps as quantum gates, rather than having the quantum steps run on quantum hardware.

    \subsection{Initialization}
        The first stage of a QGA is qubit initialization (Tab. \ref{tab:init}). This step is random when the optimization problem is encoded entirely in the fitness function. A problem-specific map is advised when the fitness function allows an optimization of the initial state to speed up convergence, as is the case of the eigenvalue problem \cite{tilly2022variational} and subprocesses of fuzzy neural network \cite{su2020prediction}. In these instances, the initial state is deemed to encode part of the problem. As an illustration, although the qubits are initialized randomly by Amal \emph{et al.} \cite{amal2022quantum}, the Thomson problem admits a mapping from the Cartesian coordinates to density matrices for each $i$th electron on the unit sphere given by the Bloch sphere and defined by Eq. \ref{eq:initthomson}, where $\hat{n}_i = [x_i,y_i,z_i]$ is the coordinate of each electron, $\bm{\sigma}=[X, \, Y, \,Z]$ is the vector of Pauli gates, $I$ is the identity matrix and $\mathrm{Tr}$ is the trace operation.

        \begin{align} \label{eq:initthomson}
            \begin{split}
            \rho_i &= \frac{1}{2}(I+\hat{n}_i\cdot \bm{\sigma} ) \\
            &\updownarrow \\
            [x_i,y_i,z_i] &= [\mathrm{Tr}(\rho_i \sigma_1), \, \mathrm{Tr}(\rho_i \sigma_2), \, \mathrm{Tr}(\rho_i \sigma_3) ]
            \end{split}
        \end{align}

        Phase initialization is exemplified in Tab. \ref{tab:init}, where $P(\theta_k)$ is the phase gate (preceded by a Hadamard gate) on the $k$th qubit $\ket{0_k}$, and $\theta_k$ is usually a random angle. During the Hamiltonian initialization, the Ansatz may also be random, although it is usually a pre-determined gate obtained during the Hamiltonian preparation. In a similar way to a Variational Quantum Eigensolver, the Hamiltonian is expressed as a linear combination of Pauli words, with one of the terms of the sum being represented in Tab. \ref{tab:init} by $U_k$.

        \begin{table}[h]
            \centering
            \caption{Example circuits of different initialization procedures for QGAs.}
            \label{tab:init}
            \begin{tabular}{p{1.7cm}lp{0.9cm}} \hline
               Initialization & Quantum circuit & Usage \\ \hline
                 Phase initialization &  \initphase & \cite{su2020prediction}
                 \\ \hline 
                 Hamiltonian initialization & \initham & \cite{tilly2022variational, ibarrondo2023quantum}
                 \\ \hline
            \end{tabular}

        \end{table}
        
        Two types of hierarchical encoding have been described in the literature: (i) the qubit as a gene, with a collection of qubits as a chromosome (or individual) and a collection of chromosomes as a population \cite{amal2022quantum}, (ii) a state from the superposition of states of the quantum register as an individual \cite{alvarez2014biomimetic, ibarrondo2022quantum, ibarrondo2023quantum}. More broadly, an arbitrarily large set of qubits and quantum registers can be encoded as individuals to avail a more complex evolutionary dynamics \cite{ibarrondo2023quantum}.

    
    \subsection{Fitness computation}
        The fitness score is computed after initialization and at the start of each iteration step. If implemented as a classical algorithm, it occurs after measurement of the quantum register and application of the fitness function over the output state.

\begin{figure}[h]
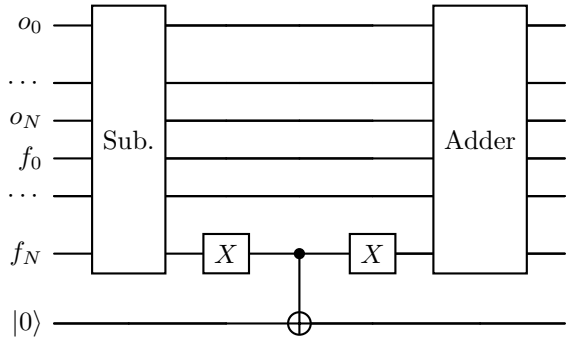

            \centering
            \foracle
            \caption{Fitness oracle in a Grover-search-based QGA computing fitness values from the individuals $o_i$ into the fitness qubits $f_i$. The Subtractor and Adder gates are explained in \cite{udrescu2006implementing}. Adapted from \cite{udrescu2006implementing}.}
            \label{fig:fitness}
        \end{figure}

        If implemented as a quantum gate, an auxiliary register is required to store the fitness of each individual \cite{udrescu2006implementing}. This type of implementation (Fig. \ref{fig:fitness}) halves the size of the population compared to a hybrid procedure with qubit individuals, but is an essential step toward a \textit{fully quantum} QGA. The oracle in Fig. \ref{fig:fitness} illustrates how the map from individual qubits $o_i$ to fitness qubits $f_i$ occurs, with an additional qubit at the bottom serving as the oracle workspace.

    \subsection{Fitness selection}
        Fitness selection is the process of sorting individuals by fitness to extract a subset of the best fit \cite{lahoz2016quantum}. Since other QGA steps have constant complexity, fitness selection is the slowest part of a QGA \cite{malossini2008quantum, saitoh2014quantum}.
        
        If implemented classically, given the fitness score of all individuals, the selection is a trivial sorting algorithm with complexity $O(n\log n)$, where $n$ is the population size \cite{malossini2008quantum}. In contrast, the quantum implementation as Grover's Search has complexity $O(1)$. It was found by Malossini \emph{et al.} \cite{malossini2008quantum} that, since Grover's Search is best suited for unstructured data, the quantum speedup is more significant when Grover's Search is implemented with a fitness function that changes at each iteration step.

        This analysis set a new milestone in QGAs by categorizing two distinct approaches based on the data structure: (i) unstructured data achieves superior time advantage in a Grover-based QGA with dynamic fitness function, while (ii) structured data can be used with either dynamic or static fitness functions with no clear difference in efficiency between them \cite{malossini2008quantum}.

        The Quantum Genetic Sampling (QGS) generalizes Grover's Search without the requirement of a fitness register. It is based on the encoding of the fitness function as a quantum gate that acts directly on the individuals, increasing the amplitude of the best-fit and leading to the measurement of the most probable state at the end of each iteration. If performed for a number of runs greater than one at each iteration, it allows the user to select a set of best-fit individuals to compose the mating pool for the next generation. \cite{acampora2022using}. 
        
        Other sorting algorithms can be easily implemented as quantum gates as an alternative to Grover's search and its variants. Bubble Sort is such an example, making use of control gates for comparison and phase inversion. In the Ibarrondo-Gatti-Sanz implementation, six ancilla qubits $\ket{a_m}$ are used to store comparisons of each of four registers $r$ \cite{codish2019sorting, ibarrondo2023quantum}.

        The Adaptive Rotation Angle is explicitly designed to leverage hybrid routines, as the rotation angles $\theta_k$ are updated after each iteration for faster convergence \cite{ballinas2023hybrid}. A simplified version of the circuits is shown in Tab. \ref{tab:fs}.

    \begin{table*}[h]
            \centering
            \caption{Example circuits of  different fitness selection procedures for a QGA.}
            \label{tab:fs}
            \begin{tabular}{p{6.1cm}ll} \hline
                Fitness selection & Quantum circuit  & Usage \\ \hline
                 Grover's Search &  \fsgrover & \cite{udrescu2006implementing, malossini2008quantum}
                 \\ \hline 
                 Quantum Genetic Sampling & \fsgeneticsampling & \cite{saitoh2014quantum, acampora2022using}
                 \\ \hline 
                 Bubble Sort & \fsbubblesort  & \cite{codish2019sorting, ibarrondo2023quantum}
                 \\ \hline
                 Adaptive Rotation Angle & \fsadaptive  & \cite{lahoz2016quantum, ballinas2023hybrid}
                 \\ \hline
            \end{tabular}
        \end{table*}

        \subsection{Quantum Crossover}
            After the individuals have been sorted, the next step is the reproduction of the best-fit, also known as crossover. Since information cannot be deleted or cloned in a unitary system, the quantum crossover is implemented by exchanging part of the information of the 50\% best-fit individuals with the 50\% least-fit ones. This is done by either (i) pseudo-cloning \cite{alvarez2014biomimetic, lahoz2016quantum} or (ii) using a SWAP gate to directly swap amplitudes between the two groups \cite{ibarrondo2023quantum}.

            The No-Cloning theorem limits the fidelity of the quantum cloning machine for the reproduction of individuals in a quantum circuit, as it can never be unit for arbitrary states. Hence, the name "pseudo-cloning" is used to refer to a set of imperfect cloning machines \cite{alvarez2014biomimetic}, drawing a distant analogy to the meiotic reproduction in cellular organisms.

            For instance, after the works of Alvarez-Rodriguez et al. \cite{alvarez2014biomimetic}, Ibarrondo, Gatti and Sanz \cite{ibarrondo2022quantum, ibarrondo2023quantum}, pseudo-cloning became a signature part of QGA research, benefiting both the areas of quantum optimization and fundamental studies in quantum cloning machines. The two main examples of pseudo-cloners are the Biomimetic Cloning of Quantum Observables (BCQO) \cite{alvarez2014biomimetic} and the Bu\v{z}ek-Hillery Universal Quantum Cloning Machine (UQCM) \cite{buvzek1996quantum, werner1998optimal}, illustrated in Tab. \ref{tab:pseudocloning}. The sub-gates of UQCM were not explicitly shown due to the length of the circuit.

            The convergence speed and fidelity results allow us to propose use cases for BCQO and UQCM. The faster performance of BCQO \cite{ibarrondo2023quantum} makes it a suitable approach for search space reduction and molecular energy ranking \cite{lima2025}. In contrast, the higher fidelity of UQCM \cite{ibarrondo2023quantum} makes it better suited for problems that require high precision, such as molecular eigensolving.
            
            Unlike the classical crossover, in pseudo-cloning-based quantum crossovers, the parents do not co-exist with their offspring, as the whole register becomes a new offspring \cite{alvarez2014biomimetic}.

           \begin{table*}[h]
                \centering
                \caption{Example circuits of different quantum crossover routines.}
                \label{tab:pseudocloning}
                \begin{tabular}{p{9cm}p{6.2cm}l} \hline
                    Quantum crossover & Quantum circuit & Usage \\ \hline \vspace{-0.8cm}
                     Biomimetic Cloning of Quantum Observables (BCQO) & \crossovercnot 
                     & \cite{alvarez2014biomimetic, ibarrondo2023quantum}\\ \hline \vspace{-0.8cm}
                      Bu\v{z}ek-Hillery Universal Quantum Cloning Machine (UQCM) & \crossoveruqcm & \cite{ibarrondo2023quantum}
                     \\ \hline  
                \end{tabular}

            \end{table*}
            
        \subsection{Quantum Mutation}
            The quantum mutation step is the application of gates that produce random changes in the amplitudes of the individuals. Three examples have been described in the literature (Tab. \ref{tab:mutations}): (i) a phase gate with a random phase $\theta$ on each qubit, (ii) random placement of a random Pauli matrix or rotation gate $U$, and (iii) the multi-control multi-target gate. Its main purpose is to prevent the algorithm from being trapped in local optima \cite{huang2021optimal}.

            The most widely used quantum mutation is the application of a random selection of $X$, $Y$ or $Z$ gates over a random selection of qubits. This is represented in Tab. \ref{tab:mutations} by raising 
            $U \in \{X,\,Y,\,Z\}$ to a random integer power $w_k\in \{ 0,1\}$ at the $k$th qubit.

            Although mutation helps the algorithm explore a larger space and avoid being trapped in local optima, if the probabilities of mutation are kept constant, it may lead to noisy or slow convergence \cite{ibarrondo2022quantum}. This issue is amended by finely tuning the scale of the random rotation angle (in case of a rotation gate as mutation) at each generation to accompany the increasing accuracy near convergence \cite{amal2022quantum}.

            In contrast, HQGA implementations in which random applications of Pauli gates have negligible effect on fidelity were also reported \cite{ibarrondo2023quantum}. These examples suggest that quantum genetic algorithms can be designed to be resilient to noise, in the analogy between mutation and quantum noise.
            
            \begin{table}[h]
                \centering
                \caption{Example circuits of different quantum mutations.}
                \label{tab:mutations}
                \begin{tabular}{p{3.2cm}lp{0.7cm}} \hline
                    Quantum mutation & Quantum circuit & Usage \\ \hline
                     Random-phase phase gate &  \pmutation
                     & \cite{amal2022quantum} \\ \hline 
                     Random-location random gate & \umutation & \cite{ibarrondo2023quantum}
                     \\ \hline  \vspace{-1.2cm}
                     Multi-control multi-target CNOT & \cnotmutation & \cite{saitoh2014quantum}
                     \\ \hline 
                \end{tabular}

            \end{table}
   \section{Implementations} \label{s:implementations}
        The term "Quantum Genetic Algorithm" has been used in the literature to refer to a range of different concepts, including classical genetic algorithm for quantum systems \cite{grigorenko2000evolutionary, grigorenko2002calculation}, quantum-inspired genetic algorithm \cite{roy2014optimization, suo2020quantum, yan2021analysis}, quantum-classical hybrid genetic algorithm \cite{ibarrondo2022quantum}, and unsupervised quantum genetic algorithm \cite{ardelean2022graph}. This is evidenced by the recurring need to emphasize the distinction between these terminologies in recent publications, to avoid ambiguity \cite{ibarrondo2023quantum}.
        As seen in the previous section, the transition from a hybrid to a fully unsupervised quantum genetic algorithm is marked by a high degree of customizability, while always keeping the mutation and crossover steps as quantum gates. The representation of the fitness calculation as a quantum gate is a milestone towards a fully quantum (i.e., unsupervised) QGA, as proposed by Rylander, Soule and Foster \cite{rylander2001quantum}. Moreover, the fitness selection as a quantum gate fulfills the final requirement for a QGA to run with a single measurement, without the need of an optimization loop \cite{ardelean2022graph}. 
        In this section, we describe these milestones while highlighting their advantages and disadvantages.

        \subsection{General Hybrid QGA} %
        A General Hybrid Quantum Genetic Algorithm (HQGA) integrates quantum computing primitives with classical evolutionary optimization, deploying quantum gates for core operations such as qubit initialization, crossover, and mutation while classically managing fitness evaluation and selection processes \cite{ibarrondo2022quantum, ballinas2023hybrid, ardelean2022graph}. Unlike fully quantum models, hybrid QGAs leverage quantum parallelism selectively to minimize resource demands on noisy intermediate-scale quantum (NISQ) devices \cite{ballinas2023hybrid}. This integration bridges classical scalability with the potential speedup provided by quantum computing, making hybrid QGAs particularly suitable for combinatorial optimization and parameter tuning \cite{ibarrondo2022quantum,ibarrondo2023quantum}.
        The HQGA workflow (Fig. \ref{f:qgaall}-a) consists of the following primary quantum operations:
        
        \paragraph{Qubit Initialization} Qubits are prepared in superposition via Hadamard gates (Eq. \ref{e:H}), enabling simultaneous representation and exploration of all possible states \cite{ibarrondo2022quantum, ardelean2022graph}.
        \begin{align} \label{e:H}
        H\ket{0} = \frac{\ket{0} + \ket{1}}{2}
        \end{align}
         For example, Ballinas \& Montiel (2023) demonstrated that adaptive Hadamard initialization enhances population diversity by approximately 32\% in Knapsack optimization problems \cite{ballinas2023hybrid}, in agreement with previous studies by Xiong \emph{et al.} stating that dynamic rotation gates improve convergence speed in Quantum-Inspired Evolutionary Algorithms \cite{xiong2018quantum}.

        \paragraph{Quantum Crossover} Quantum crossover operations utilize entanglement, typically through Controlled-NOT (CNOT) gates (Eq. \ref{e:cnot}), to exchange genetic information between parent chromosomes \cite{ardelean2022graph}.
        \begin{align} \label{e:cnot}
            \mathrm{CNOT}\ket{q_{1}q_{2}} = \ket{q_1} \otimes \ket{q_{1} \oplus q_2}
        \end{align}
        This method significantly enhances efficiency. For example, Udrescu \emph{et al.} (2022) applied entangled crossover in graph coloring tasks, achieving a notable 45\% reduction in memory usage \cite{ardelean2022graph}.
        
        \paragraph{Quantum Mutation} Quantum mutation is implemented through rotation gates $R_y(\theta)$, which probabilistically alter qubit states: 
        \begin{align}
            R_y(\theta) = \begin{pmatrix}
                \cos(\theta/2) & -\sin(\theta/2)\\
                \sin(\theta/2)&\cos(\theta/2)
            \end{pmatrix}
        \end{align}
        Here, the rotation angle $\theta$ is dynamically tuned to maintain a balance between exploration and exploitation \cite{ballinas2023hybrid}. Zhang \emph{et al.} (2023) successfully coupled rotation gates with deep learning in video tracking, significantly improving robustness against noise \cite{zhang2023quantum}.

        HQGAs have already shown substantial practical advantages, such as a 40\% faster convergence rate in Knapsack problems via parallel quantum crossover and mutation, a 22\% reduction in energy consumption for MIMO optimization using shallow, decoherence-resistant circuits, and approximately 50\% shorter runtimes when heavy fitness evaluations are classically managed, as demonstrated in building envelope design \cite{wang2021design,almasaoodi2023optimizing,ballinas2023hybrid}. They also integrate seamlessly into established engineering workflows like IoT antenna miniaturization and fog-cloud scheduling \cite{bichara2023quantum,belmahdi2022sqga}. However, classical bottlenecks in fitness evaluation and selection can significantly limit quantum speedup potential, particularly for large-scale instances \cite{ibarrondo2022quantum, ardelean2022graph}. Additionally, precise parameter tuning for quantum rotations is challenging, and qubit decoherence in deeper quantum circuits restricts current device scalability \cite{ibarrondo2023quantum,ballinas2023hybrid}. Such limitations may amplify latency; for instance, frequent quantum-classical data exchanges during molecular optimization have resulted in runtime increases of approximately 15\% \cite{ibarrondo2022quantum}.

        \subsection{Rylander-Soule-Foster Hybrid QGA} 
\begin{figure}[ht]
            \centering
            \includegraphics[width=0.7\linewidth]{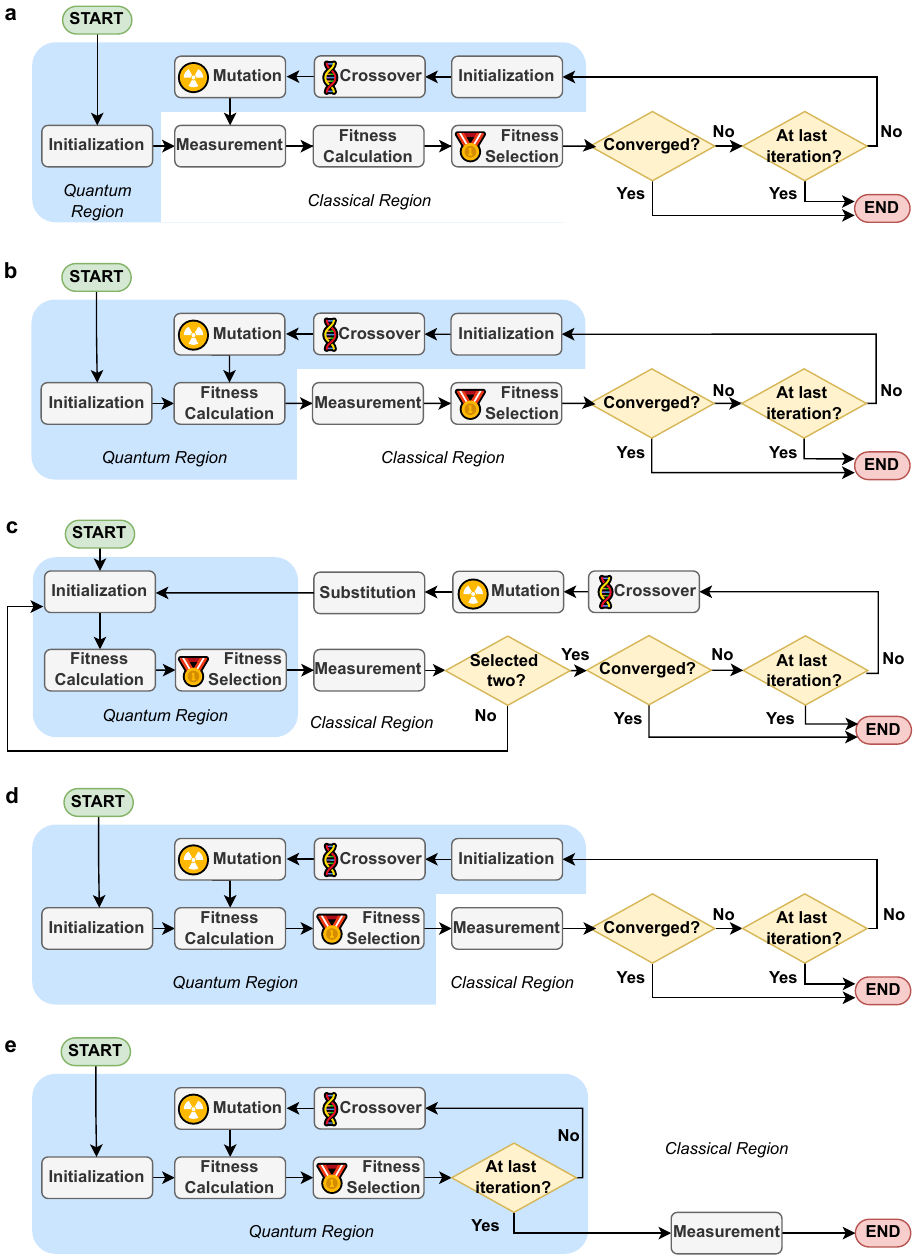}
            \caption{Most common layouts of Quantum Genetic Algorithms from hybrid to unsupervised. (a) A General Hybrid QGA shows quantum crossover and mutation, but classical fitness calculation and selection. (b) A Rylander-Soule-Foster Hybrid QGA \cite{rylander2001quantum} adds a quantum register to store the fitness calculation. (c) A Quantum Genetic Optimization Algorithm only implements fitness calculation and selection as quantum routines, leaving the rest classical \cite{malossini2008quantum}. (d) A Reduced Quantum Genetic Algorithm has all steps as quantum gates, but is iterated classically \cite{udrescu2006implementing} (e) An hypothetical Unsupervised QGA has all steps as quantum gates, a pre-defined number of iterations (powers of the circuit gate), and only one measurement step.}
            \label{f:qgaall}
        \end{figure}

            Inclusion of the fitness score calculation in the quantum circuit (Fig. \ref{f:qgaall}-b) requires dedicating qubits for storage. Rylander, Soule and Foster \cite{rylander2001quantum} envisioned this step by designing a quantum circuit with a separate quantum register for the population and another one for the fitness. The qubits of the population register are entangled with those of the fitness register, resulting in an automatic update of the fitness by any operation over the individuals. 

            In this model, the quantum individual is defined as equivalent to a quantum state, i.e., a superposition of states, with each state representing a classical individual. The definition of which specific superposition is a relevant quantum individual to undergo operations is a convenience of how to implement the model, with no known study exploring the subject.

            Naturally, the best-fit individual is the one whose fitness qubit has the highest probability. By measuring only the fitness qubits, the population is reduced to a superposition of individuals of the same fitness.

            Unlike the HQGA, the Rylander-Soule-Foster model \cite{rylander2001quantum} explicitly exploits entanglement in order to obtain a speedup advantage. Although it is mentioned by renowned authors \cite{ibarrondo2023quantum, ardelean2022graph} as a contrast to hybrid models, the routine is still supervised and iterated by a classical computer, as most quantum-hybrid optimizers.

        \subsection{Reduced QGA} 
            The final landmark towards a \textit{fully} quantum genetic algorithm (Fig. \ref{f:qgaall}-c) is the implementation of Grover's Search \cite{grover1997quantum} by Udrescu \emph{et al.} in \cite{udrescu2006implementing, ardelean2022graph}. Their Reduced QGA (RQGA) restricted the role of the classical computer to the measurement and iteration steps, with all other QGA operations being quantum.
            With fitness marking as a quantum gate, the superposition is only destroyed at the measurement, after the amplitude amplification of the best-fit individual provided by Grover's Search. Subsequently, the measured result is used to update the amplification threshold of Grover's Search at each classical iteration \cite{ardelean2022graph}.

            Udrescu's RQGA does not include a quantum procedure to update Grover's search parameters at each generation. This led Saitoh to propose an extension to Grover's search, named Quantum Genetic Sampling \cite{saitoh2014quantum}. At this stage, a potential non-iterative algorithm with a single measurement after a predetermined number of generations is expected to be less efficient than a hybrid version due to the lack of adaptability to the new best-fit value in each generation \cite{ardelean2022graph}.

\section{Designing QGAs for physical systems} \label{s:tailoring}

    The fitness function of the problem evaluates each individual in the population, steering the search toward chromosomes that best satisfy the problem requirements. In each generation, the two fittest individuals are selected for crossover and mutation, producing two offspring that replace two randomly chosen population members. This cycle repeats until either the population's aggregate fitness surpasses a prescribed threshold or a fixed number of genetic iterations has been executed.

    In the quantum setting, any classical reversible circuit that computes a fitness function $F(j) = F_j$, where $j \in \{0, \dots, N-1\}$ are the elements of the population, can be promoted to a quantum fitness evaluation operator $U_F$. Encoding the elements in quantum binary, the superposition of all $N$ individuals in the population is denoted by

    \begin{align}
        \ket{\Psi} = \frac{1}{\sqrt{N}} \sum_{j=0}^{N-1} \ket{j}
    \end{align}

    Hence, using $U_f$ only once entangles each basis state $\ket{j}$ with its corresponding fitness value, so that the fitness $\{F_j |j = 0,\, \dots, \, N-1\}$ of the entire population is evaluated in a single quantum operation. In contrast, the classical procedure requires $N$ fitness evaluations. The fitness value of each population element reflects the quality of the characteristics encoded by individual $j$.
    Because the shape of the fitness landscape heavily influences convergence, the original cost (objective) function should be transformed into a fitness function that accelerates progress, avoids local minima, and reduces noise \cite{ibarrondo2022quantum}. This section describes the lookup tables used to map bit-strings to real numbers, the most used objective functions to compute the individual's fitness in Genetic and Quantum Genetic Algorithms, and their usage context. The functions are classified into three categories:  benchmark polynomials,  statistical functions, and physical functions.

    \subsection{Benchmark Functions}
        Symmetric power and sinusoidal functions are ideal test beds for determining the general behavior of genetic algorithms for different levels of complexity \cite{li2020quantum}. For example, the Sphere function (Eq. \ref{e:sphere}) is perhaps the simplest minimization task, with a global minimum at the origin. In this function, a global minimizer has a single radial direction to follow for any range of the search space. The Quartic function and similar multivariate monomials offer similar shapes to test simple acceleration subroutines due to the power descent towards the global minimum \cite{li2020quantum}.

        The inclusion of periodic oscillations gives rise to infinitely many local minima, significantly increasing the challenge for optimization algorithms. For simplicity, we name two categories: radial and plane. In the case of radial oscillations, high concentric barriers separate local maxima and local minima, testing the resistance of the algorithm to local-optimum traps. The most used examples are the Salomon function \cite{li2020quantum}. Oscillations on the plane generate a pattern of periodically distributed peaks and wells that will test the ability of the algorithm to navigate guided by an average of past best-fit individuals rather than exclusively by the last best-fit individual, in addition to converging to a value with minimal noise. Illustrative examples are the Rastrigin function and the Penalized function \cite{li2020quantum}.

        \begin{align} \label{e:sphere}
            f(\mathbf{x}) = \sum_{i=1}^{N} x_{i}^2
        \end{align}

        Moreover, custom distances based on polynomials, exponentials and sinusoids are also used as adaptations of the above-mentioned functions to suit various problems \cite{lin2021intelligent}.
        
    \subsection{Error Functions} 
        Error metrics are the most widely used objective functions in quantum genetic algorithms, as they constitute an explicit encoding of a minimization problem for an arbitrary problem. The fitness as a property to be maximized is then interpreted as its inverse \cite{su2020prediction}. The choice of error function from absolute error, mean squared error, or other custom distance may affect the convergence speed due to the convergence threshold \cite{su2020prediction}, but there is no consensus on which function will perform better for a general case. Nonetheless, the mean squared error is the most commonly used error function in fitness functions \cite{pan2021self, huang2020directional, su2020prediction, ge2019improved}, followed by the absolute error \cite{huang2021optimal, zhang2020application, zhang2015screw}.

        Furthermore, few studies use the signal-to-noise ratio as a function of the mean squared error, especially in applications to image processing \cite{zhang2019adaptive, wang2020adaptive}. In this case, the signal-to-noise ratio is a proper fitness function to be maximized.

    \subsection{Physical Functions}
        Functions that explicitly represent physical systems in terms of energy and have been described in QGAs include potential energy \cite{thomson1904xxiv, amal2022quantum} and total energy as the Hamiltonian of the \ce{H2} molecule \cite{tilly2022variational}.

        In addition to functions that map post-measurement bit-strings to real fitness values, fitness functions are commonly encoded in QGAs as inverse cost functions in a problem Hamiltonian. In this case, the goal of optimization is to evolve the population toward the ground state \cite{ibarrondo2023quantum}. Combinatorial optimization problems are naturally encoded as Ising Hamiltonians \cite{lucas2014ising}, and molecular ground state estimation is formulated in terms of an electronic Hamiltonian \cite{tilly2022variational}. 

        As demonstrated by \cite{amal2022quantum}, combinatorial problems that are described by a potential also have a HQGA representation. Potential energy (Eq. \ref{e:potential}) is largely used in variants of the Thomson problem \cite{thomson1904xxiv, amal2022quantum} as a fitness function to be minimized. In this problem, the coordinates of electrons on a sphere are mapped to qubits on a Bloch sphere, and the problem is solved when the potential energy of the system of particles is minimized.

        \begin{align} \label{e:potential}
            F(\hat{r}_1, \hat{r}_2, \cdots, \hat{r}_N) 
            = \sum_{s<s^\prime} 
            \frac{1}{|\hat{r}_s - \hat{r}_{s^{\prime} }| }
        \end{align}

        Since the physical information about the system is encoded in the initialized coordinates and in the potential energy as a fitness function, this problem can easily be adapted to represent other configuration problems by a simple change of potential, such as viral morphology \cite{caspar1962physical}, and molecular geometry optimization \cite{szalay2015tensor}.

    From the vast applications of QGA in Sec. \ref{sec:advantages}, two of them have the energy of a physical system as an explicit fitness function. The Thomson problem uses potential energy while the Eigenvalue problem of a molecule uses a Hamiltonian representing the total energy of the system.

    \subsection{Scalar potentials} 

        The problem of finding the configuration of electrons around a sphere that minimizes the energy of the system is known as the Thomson problem \cite{thomson1904xxiv}, with applications in molecular geometry and electronic structure determination \cite{amal2022quantum}. Its representation as a QGA problem involves mapping the possible solutions of sequences of points on a unit sphere as sequences of points on the Bloch sphere, according to Eq. \ref{e:bloch} for $x_k$ as the spatial coordinates, $\rho$ the density matrix of the state, and $\sigma_k$ as the Pauli matrices from $k=1$ to $k=3$ \cite{amal2022quantum}. 
        
        \begin{align} \label{e:bloch}
            x_k = \mathrm{Tr} [ \rho \sigma_k ]
        \end{align}
        
        Each qubit is understood as a gene, whereas a sequence of qubits is a chromosome. The fitness function to be minimized is the potential energy of the system of particles, according to Eq. \ref{e:potential}, where $\mathbf{r}_s$ is the coordinate vector of the $s$th particle.
    

    \subsection{Matrix Hamiltonians} 

        Ground state estimation is an energy minimization problem given by Eq. \ref{e:energymin}, where $H$ is the problem Hamiltonian, and $\rho$ is the density matrix representing a state. When $\rho$ is the ground state, the average energy $\braket{E}_\rho$ is the smallest eigenvalue of the problem Hamiltonian \cite{ibarrondo2022quantum}.
        
        \begin{align} \label{e:energymin}
            \braket{E}_\rho = \mathrm{Tr} [H \rho]
        \end{align}

        Rather than minimizing only the average energy, in order to obtain experimentally relevant results, the fidelity $f$ (Eq. \ref{e:fidelity}) should be maximized \cite{ibarrondo2022quantum}.

        \begin{align} \label{e:fidelity}
            f(\rho, \ket{u_1} ) = 
            \bra{u_1} \rho\ket{u_1} 
        \end{align}

        In the most extensive application of QGAs to eigensolve Hamiltonians, Ibarrondo \emph{et al.} used two approximate quantum cloning machines as the crossover step. The biomimetic cloning of quantum observable (BCQO) preserves the statistics of the state after cloning, but its fidelity is variable and input-dependent \cite{alvarez2014biomimetic}. The second one, the Buzek-Hillery universal quantum cloning machine (UQCM), clones the fidelity perfectly, but the states are not identical clones \cite{buvzek1996quantum}. As noted by Ibarrondo \emph{et al.}, UQCM and fidelity preservation play a greater role in increasing both convergence speed and fidelity \cite{ibarrondo2022quantum}, being the most efficient choice for QGA eigensolvers.
    
        \begin{table*}[h]
            \centering
            \caption{Fitness functions and their applications.}
            \label{tab:ffa}
            \begin{tabular}{llllll}
            \hline
                 Fitness function & Name & Image & Application & Year & Ref  
                 \\ \hline
                 $f=1-\frac{1}{s}\sum_{p=1}^{s}\left| \frac{y^{\prime}_p-t_p}{t_p}\right|$ & Absolute error & $[0,1]$&Neural network & 2015 & \cite{zhang2015screw}
                 \\$f = \frac{1}{N}\left(\sum_{i=1}^{N} x_i^2 - Nx_\mathrm{mean}^2\right)$ & Mean square error & $(-\infty, \infty)$& Visual tracking & 2019 & \cite{ge2019improved}               
                 \\ $f=x^2+y^2$ & Sphere function & $[0,\infty]$ & Bench-marking &2020 & \cite{li2020quantum}
                 \\  $f=-\sum_{i=0}^{2^N -1} p(m_i) \log_2 p(m_i)$ & Information entropy & $[0,\infty)$ & Image encryption & $2020$ & \cite{cheng2020novel}
                 \\ $f 
            = \sum_{s<s^{\prime}} 
            \frac{1}{|\mathbf{r}_s - \mathbf{r}_{s^\prime}| }$ & Potential energy &  $[0,\infty)$ & Geometry optimization & 2022 & \cite{thomson1904xxiv}
                 \\ $f=\Delta \mathrm{Tr}[H\rho]$ & Average energy change & $(-\infty, \infty)$ &Molecular eigensolving & 2022 & \cite{tilly2022variational}
                 \\ \hline
            \end{tabular}

        \end{table*}


\section{Conclusion} \label{s:conclusion}

In this survey, we consolidated three decades of research in Quantum Genetic Algorithms into a coherent framework that connects algorithm design, quantum advantage and applications. We provided a roadmap of QGA research, and systematically classified QGA architectures, from hybrid models to Reduced QGAs and fully unsupervised quantum genetic routines.

By exemplifying different approaches to encode initialization, fitness computation, selection, crossover, and mutation as quantum gates, we established a clear distinction between genuinely quantum genetic algorithms and quantum-inspired GAs. While the crossover step makes quantum advantage an emergent phenomenon in QGAs, we identified the selection step as the dominant source of quantum speedup over the closest-related classical GAs, with Grover Search in RQGAs as the leading routine.

Finally, we discussed how the Thomson problem and the eigenvalue problem facilitate the study of physical systems governed by scalar potentials and matrix Hamiltonians, in addition to having highlighted topics of future research, such as the implementation of an unsupervised QGA and generalizations of the Thomson problem.


\section*{Aknowledgements} 
    This work was funded by Qatar Center for Quantum Computing and Hamad Bin Khalifa University.


\section*{Author contributions}

Dennis Lima: Conceptualization, data curation, investigation, visualization, writing -- original draft. Rakesh Saini: Conceptualization, data curation, investigation, visualization, writing -- original draft. Saif Al-Kuwari: Funding acquisition, supervision, writing -- review and editing.


\bibliographystyle{unsrt}
\bibliography{main}

\end{document}